\documentclass[journal]{IEEEtran}

\usepackage{float}
\usepackage{graphicx}
\usepackage{color}
\usepackage[square, comma, sort&compress, numbers]{natbib}
\usepackage{hyperref}
\usepackage{subfigure}
\usepackage{amsmath}
\usepackage{clrscode}
\usepackage{algorithm}
\usepackage{algorithmic}
\usepackage{multirow}
\usepackage{multicol}
\usepackage{mathtools}
\usepackage{amsfonts}
\usepackage{array}
\usepackage{amsthm}
\usepackage{amssymb}
\usepackage{makecell, multirow}
\usepackage{tikz}
\usepackage{pgfplots}
\usepackage{booktabs}
\usepackage{optidef}
\usepgfplotslibrary{units}
\usetikzlibrary{spy,backgrounds}
\usepackage{pgfplotstable}

\begin{document}
%
\title{Reinforcement Learning for Robust Header Compression under Model Uncertainty}

\author{
 Shusen Jing, ~\IEEEmembership{Member,~IEEE},  Songyang Zhang, ~\IEEEmembership{Member,~IEEE} and Zhi Ding, ~\IEEEmembership{Fellow,~IEEE} 

\thanks{All of the authors are with the Department of Electrical and Computer Engineering, University of California, Davis, 95616, e-mail: shjing@ucdavis.edu, sydzhang@ucdavis.edu and zding@ucdavis.edu.}
\thanks{Copyright (c) 2020 IEEE. Personal use of this material is permitted. However, permission to use this material for any other purposes must be obtained from the IEEE by sending a request to pubs-permissions@ieee.org.}
}

\markboth{IEEE TRANSACTIONS ON WIRELESS COMMUNICATION, VOL. xx, NO. xx, July 2023}
{Shell \MakeLowercase{\textit{et al.}}: Bare Demo of IEEEtran.cls for IEEE Journals}

\maketitle

\begin{abstract}
Robust header compression (ROHC), critically positioned between the network and the MAC layers, plays an important role in modern wireless communication systems for improving data efficiency.
This work investigates bi-directional ROHC (BD-ROHC) integrated with a novel architecture of reinforcement learning (RL). We formulate a partially observable \emph{Markov} decision process (POMDP), in which agent is the compressor, and the environment consists of the decompressor, channel and header source. Our work adopts the well-known deep Q-network (DQN), which takes the history of actions and observations as inputs, and outputs the Q-values of corresponding actions. Compared with the ideal dynamic programming (DP) proposed in the existing works, 
our method is scalable to the state, action and observation spaces. In contrast, DP often suffers from formidable computational complexity when the number of states becomes large due to long decompressor feedback delay and complex channel models.
In addition, our method does not require prior knowledge of the transition dynamics and accurate observation dependency of the model, which are often not available in many practical applications.
\end{abstract}


\IEEEpeerreviewmaketitle

\section{Introduction}


Advancements in recent communication generations have greatly enhanced bandwidth efficiency through technologies at the PHY/MAC layer, reaching performance levels close to their limits \cite{phy}. Consequently, little room is left for further improvement at PHY/MAC layer. With the widespread adoption of Internet Protocols (IP) in numerous applications and services, the move towards all-IP packet-switched architectures in wireless network infrastructures has become prominent \cite{allip}. Future improvements in wireless networks should not only concentrate on MAC and PHY layer techniques, but also encompass a greater focus on optimizing IP-based protocol stacks across wireless infrastructures.


An IP packet consists of a header and a payload, where the header contains essential system information such as version, time to live, and IP addresses. In certain applications, such as Voice-over-Internet-Protocol (VoIP) and Internet-of-Things (IoT), the header can be comparable to, or even larger in size than, the payload, which can compromise the overall data efficiency of packets transmission. To address this issue, a mechanism called robust header compression (ROHC) was developed \cite{bormann2001robust,degermark2001requirements,jonsson2007robust,jonsson2004robust}. ROHC takes advantage of the fact that many fields in the header tend to change slowly throughout the lifetime of a data flow. It selects reference values for these fields and only encodes the small deviations from these reference values in the header. By compressing the header in this manner, ROHC reduces the overhead associated with transmitting IP packets, thereby improving packet network efficiency. The adoption of ROHC has been widespread in wireless packet switch networks such as 4G-LTE \cite{3gppevolved} and 5G-NR \cite{feres2019low}, and it has a strong potential in IoT scenarios where packets with short payload are prevalent. By minimizing the header length without sacrificing important system information, ROHC enhances the efficiency of IP-based communication in various wireless networks.


Despite its widespread deployment, ROHC has not received significant attention. However, a few studies have focused on improving and analyzing the performance of ROHC. As an early work, the authors of \cite{rohc3} proposed configurations of ROHC for scenarios with scarce resources links to improve efficiency and robustness. Window-based least significant bits (W-LSB) encoding \cite{bormann2001robust} is one of the common compression method in ROHC. The authors of \cite{rohc1} studied the impact of the different channel conditions on W-LSB encoding in ROHC. It pointed out that smaller window size is preferable when channel condition is good. However, the existing works only considered memoryless channels. A more recent work \cite{rohc2} filled this gap by adopting Gilbert Elliot dynamic channel model in ROHC, and described the system behavior with mathematical models. The proposal of \cite{rohc4} leveraged hybrid ARQ (HARQ) information from PHY/MAC layer to facilitate ROHC design. More recently, the authors of \cite{wenhao1} first formalized the U-mode ROHC as a partially observable \emph{Markov} decision process (POMDP), in which trans-layer information, including HARQ, is used as partial observation to support the decision-making of compressor. A subsequent work \cite{wenhao2} proposed the bi-directional ROHC (BD-ROHC), in which the compressor may require feedback from the decompressor. It formulated BD-ROHC as a POMDP and proposed the optimal solution using  dynamic programming (DP).

Although the optimal solution provided by DP \cite{wenhao2} for the BD-ROHC, it has two practical issues. First, DP becomes computationally prohibitive when the number of POMDP states becomes larger, resulting from long feedback delay and complicated channel models. Second,  DP, together with other existing approaches, rely on transition dynamics and probabilistic of POMDP model, which is often unavailable or inaccurate in many practical applications. To address these issues, we propose a BD-ROHC design using reinforcement learning (RL). Specifically, we adopt deep Q-network (DQN) to incorporate the history of actions and observations as inputs, and generate as outputs the Q-value corresponding to each action. This RL framework enables us to handle POMDP with a vast number of states, including the cases where the state space is infinite. Moreover, DQN's training process only relies on collected episodes, eliminating the need for explicit knowledge of transition dynamics. The double DQN (DDQN) technique is further deployed to improve the convergence and stability of the learning process. Our simulation results demonstrate better transmission efficiency achieved by our proposed RL method than benchmarks under different channel models and parameter settings, without prior knowledge of model dynamics.

The rest of this paper is organized as follows: Section \ref{sec:sys} reviews the system model, including the basic functionality of compressor and decompressor. Section \ref{sec:pomdp} delivers the POMDP formulation. Section \ref{sec:dqn} provides details on the deployment of the deep Q-learning for BD-ROHC. Section \ref{exp1} demonstrates the the proposed design through simulations. Section \ref{sec:conl} finally summarizes the work.

\section{System Model}\label{sec:sys}
\begin{figure}[htbp]
\centering
\includegraphics[width=\linewidth]{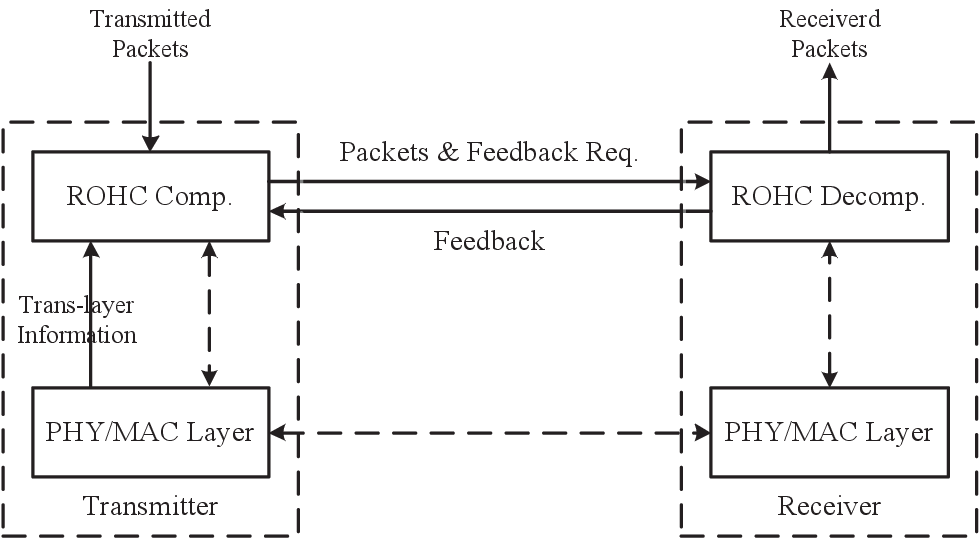}
\caption{The system diagram of BD-ROHC.}
\label{fig:general_rohc}
\end{figure}

Fig. \ref{fig:general_rohc} presents the system diagram of BD-ROHC. The compressor selects compressed headers with different lengths for the packets to be transmitted. The decompressor tries to decode headers of packets received from the channel. Decoding failures could happen due to the imperfection of channels or over-aggressive compression of the headers. The compressor uses trans-layer information from lower layer (MAC/PHY) as observations to support its decision making, such as channel quality information (CQI), hybrid ARQ (HARQ) feedback and frequency of header context initialization. In the bi-directional setting, the ROHC compressor may also request feedback from the decompressor, which can be used for decision making as well. We now discuss the components and functionalities of BD-ROHC in details.

\subsection{Headers}
Let $\alpha_C[t]$ denote the compressor's decision on the headers at the $t$-th slot. There are three choices of headers, namely initialization and refresh (IR) header, compressed header with $7$-bits CRC (CO7) and compressed header with $3$-bits CRC (CO3), denoted as $\alpha_C = 0,1,2$, respectively.

\begin{itemize}
    \item \textbf{IR header} ($\alpha_C[t] = 0$, longest with length $L_0$) is the full length header, which is used to establish header context at the decompressor. The decompressor can decode compressed header, i.e., CO7 and CO3, only if the header context has been established. 
    \item \textbf{CO3 header} ($\alpha_C[t] = 2$, shortest with length $L_2$) is the fully compressed header. It can be decoded only if the decompressor maintains a header context. Due to the imperfection of the channels, decoding failures can happen even if header context is maintained. After a few successive failures, the context will be damaged and CO3 will not be useful.
    \item \textbf{CO7 header} ($\alpha_C[t] = 1$, of medium length $L_1$) is used for repairing the damaged header context. If the decompressor successfully decode a CO7 header, the damaged header context can be repaired, after which CO3 header can be decoded again with the header context.
\end{itemize}

In general, longer headers are more likely to be successfully decoded by the decompressor, but it has low packet transmission efficiency defined as $L/(L+L_i)$, where $L$ is the length of the payload.

Note that headers are not always compressible. Whether a header is compressible depends on the header sources at the transmitter. We denote the compressibility of the header at the $t$-th slot as $ \sigma_S[t]$, and use $ \sigma_S[t]=1$ and $ \sigma_S[t]=0$ to represent compressible and uncompressible header, respectively. If a header is uncompressible, an IR or CO7 header will be required, i.e., $\alpha_C[t]=0,1$, and CO3 header will not be taken by the decompressor. We assume $ \sigma_S[t]$ evolves as a $d_S$-th order \emph{Markov} model with dynamic $\mathcal{T}_S( \sigma_S[t]| \sigma_S[t\!-\!1\!:\!t-d_S])$, which represents the probability distribution of the future header compressibility $\sigma_S[t]$, conditioned on the history of header compressibilities $\sigma_S[t\!-\!1\!:\!t-d_S]$.

\begin{table*}[htbp]
\caption{State transition of the decompressor. }
\centering
\begin{tabular}{c|c|c|c|c|c|c} 
\hline
& $0$ (FC)  & $1$ (FC) & $2$ (FC) & $\dots$ & $W$ (RC) & $W\!+\!1$ (NC)   \\ 
\hline
$0$ (FC) & $ \sigma_T\!=\!1\! \vee\! ( \sigma_S\!=\!1 \!\wedge \!\alpha_C\!\neq \!2)$  & $ \sigma_T\! =\! 0\! \wedge\!  \sigma_S\! = \!1$ &  & $\dots$ & $ \sigma_S\!=\!0\wedge\! ( \sigma_T\!=\!0 \!\wedge\! \alpha_C \!=\!2)$ &     \\ 
\hline
$1$ (FC) & $ \sigma_T\!=\!1 \!\vee\! ( \sigma_S \!=\!1 \!\wedge \!\alpha_C\!\neq\! 2)$  &  & $ \sigma_T\! =\! 0 \!\wedge\!  \sigma_S \!=\! 1$ & $\dots$ & $ \sigma_S\!=\!0\!\wedge\! ( \sigma_T\!=\!0 \!\wedge \! \alpha_C\! =\!2)$ &     \\ 
\hline
$\vdots$ & $ \sigma_T\!=\!1 \!\vee \!( \sigma_S \!=\!1 \!\wedge\! \alpha_C\!\neq \!2)$  &  &  & $\ddots$ & $ \sigma_S\!=\!0\wedge\! ( \sigma_T\!=\!0 \wedge \alpha_C \!=\!2)$ &     \\ 
\hline
$W\!\!-\!1$ (FC) & $ \sigma_T\!=\!1 \vee ( \sigma_S \!=\!1 \!\wedge\! \alpha_C\!\neq \!2)$  &  &  & $\dots$ & $ \sigma_T\! =\! 0 \vee ( \sigma_S \!=\!0 \!\wedge \!\alpha_C \!= \!2)$ &     \\
\hline
$W$ (RC) & $ \sigma_T\!=\!1 \!\wedge \!\alpha_C\!\neq \!2$  &  &  & $\dots$ & $ \sigma_T\!=\!0\! \vee\! \alpha_C\!=\!2$ &    \\ 
\hline
$W\!\!+\!1$ (NC) & $ \sigma_T\!=\!1\! \wedge\! \alpha_C\!=\!0$  &  &  & $\dots$ &   &  $ \sigma_T\!=\!0\!\vee\! \alpha_C\!\neq\! 0$  \\ 
\hline
\end{tabular}
\label{tab:sta}
\end{table*}

\subsection{Channel}
The channel between the compressor and the decompressor is not perfect. We denote the channel quality as $ \sigma_H[t]$ at the $t$-th slot, where larger $ \sigma_H[t]$ indicating better channel quality. We use $ \sigma_T[t]$ to denote the packet transmission status at the $t$-th slot, and use $ \sigma_T[t]= 1$ and $ \sigma_T[t]= 0$ to denote transmission success and failure, respectively.  When the transmission fails, the header can not be decoded by the decompressor. 

In this work, we do not make specific assumptions on channel models. Instead, we only assume the channel quality $ \sigma_H[t]$ evolves as a $d_H$-th order \emph{Markov} process through the dynamic $\mathcal{T}_H( \sigma_H[t]| \sigma_H[t\!-\!1\!:\!t\!-\!d_H])$, which is the probability distribution of future channel quality $\sigma_H[t]$, conditioned on the history of channel qualities $\sigma_H[t\!-\!1\!:\!t\!-\!d_H]$. The transmission status $ \sigma_T[t]$ depends on the channel quality $ \sigma_H[t]$ and the header type $\alpha_C[t\!-\!1]$ through the dynamic $\mathcal{T}_T( \sigma_T[t]|\alpha_C[t\!-\!1], \sigma_H[t])$, which can be explained as conditional distribution similarly.

\subsection{Trans-layer Information}
The compressor makes decisions on headers to use and on whether to send feedback requests leveraging trans-layer information, such as channel quality information (CQI), hybrid ARQ (HARQ) feedback and frequency of header context initialization. Usually these trans-layer information is managed by lower layer protocols, and not reported to higher layers. In our design, we assume the compressor extracts trans-layer information to facilitate its decision making. For simplification, we assume the compressor estimates the $d_D$-delayed channel quality $ \sigma_H[t\!-\!d_D]$ and transmission status $ \sigma_T[t\!-\!d_D]$ from the trans-layer information, which are denoted as $z_H$ and $z_T$ through the probabilistic $\mathcal{O}_H(z_H[t]| \sigma_H[t\!-\!d_D])$ and $\mathcal{O}_T(z_T[t]| \sigma_T[t\!-\!d_D])$, which can be explained as the distribution of $z_H[t]$ and $z_T[t]$ conditioned on delayed channel condition $\sigma_H[t\!-\!d_D]$ and delayed transmission status $\sigma_T[t\!-\!d_D]$, respectively. Again, we do not assume the model of this two probabilistic in this work.

\subsection{Decompressor}\label{sec:decomp}
The goal of the decompressor is to decode the headers received from the channels.  Whether the header can be successfully decoded by the decompressor depends on the transmitted header type $\alpha_C$, the compressibility of the header $ \sigma_S$ (decided by the header source at the transmitter), transmission status $ \sigma_T$, and the state of the decompressor $ \sigma_D$. 

The decompressor works as a finite state machine (FSM) with $W+2$ states, of which the state transition diagram can be found in Fig. 2 of \cite{wenhao2}. We use $ \sigma_D = W+1$ to represent the "No Context" (NC) state, $ \sigma_D = W$ to represent "Repair Context" (RC) state, $ \sigma_D=0,1,...,W-1$ to represent "Full Context" (FC) state with confidence from high to low. At the beginning, the decompressor is at NC state. It is not able to decode CO3 or CO7 header since there is no context established. When successfully decoding an IR header, the decompressor establishes a context and transits to $ \sigma_D=0$ FC state. At FC state $ \sigma_D=l$ with $l=0,1,...,W-2$, if transmission failure happens $ \sigma_T = 0$ and the header is fully compressible $ \sigma_S = 1$, the decompressor will transit to the lower level FC $ \sigma_D = l+1$ and claim a decoding failure. After $W$ such failures, it will transit to RC state. However, when the header is not fully-compressible $ \sigma_S = 0$, it will directly transit to RC state and claim decoding failure if transmission failure happens $ \sigma_T = 0$ or the header is fully compressed $\alpha_C = 2$. At RC state, unless it successfully decode an IR or CO7 packet, i.e., $ \sigma_T = 1$ and $\alpha_C <2$, it will stay in RC and claim decoding failure. It is worth noting that the decompressor successfully decode the header if and only if its state transits to $ \sigma_D=0$. The detailed state transition is shown in Table \ref{tab:sta}, where the $(i,j)$-th entry is the condition on which the $i$-th state transits to the $j$-th state. A blank entry means the transition can never happen.

\begin{figure*}[htbp]
\centering
\includegraphics[width=0.8\linewidth]{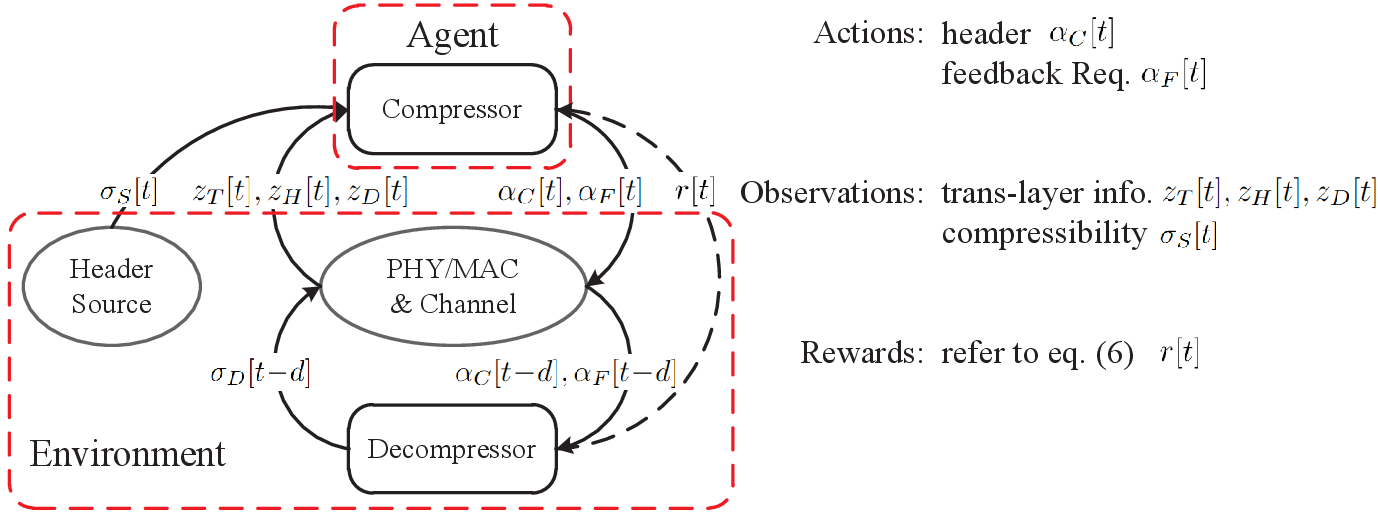}
\caption{The block diagram of BD-ROHC in the context of RL. Agent is the compressor, and the environment consists of the decompressor, channel and header source.}
\label{fig:rohcmodel}
\end{figure*}

\subsection{Compressor}
The compressor makes decisions on $\alpha_C[t]\in \{0,1,2\}$ for the $t$-th packet, i.e., deciding which one of the IR, CO7 and CO3 headers should be used. In addition, it also decides whether to request a feedback at the $t$-th slot from the decompressor to facilitate future decision making. This request allows the compressor to fully observe the decompressor state, albeit with a time delay. We use $\alpha_F[t]$ to denote the decision on whether request feedback, and denote request and not request as $\alpha_F[t]=1,0$, respectively. If the compressor send a request at the $t$-th slot, it will receive a feedback $z_D[t+d_D]= \sigma_D[t]$ at the $(t+d_D)$-th slot, indicating the state of the decompressor at the $t$-th slot.

In general, the compressor does not know the state of the decompressor. It relies on the trans-layer information $z_T[t]$ and $z_S[t]$ as partial observations to make decisions. It can also use the feedback $z_D[t] =  \sigma_D[t-d_D]$ to support the decision making  if it requested a feedback at the $(t\!-\! d_D)$-th slot. 

Note that longer headers are more likely to be successfully decoded considering the imperfection of channels and state transition of the decompressor but at the sacrifice of packet-efficiency as a result. The feedback can provide more information about the decompressor's states, but at the cost of additional communication resources. The decision making at the compressor is nontrivial considering these factors. 

\subsection{Summary of Notations}
We now using the following Table \ref{tab:1} to summarize the major notations and symbols used in our problem formulation.
\begin{table}
\caption{Summary of Notations}
\begin{tabular}{ c|c} 
\hline
Variable & Description \\
\hline
$\alpha_C$ & \makecell[l]{Compressor's decision on headers: It takes value $0,1$ and $2$ \\ for IR, CO7 and CO3 header, respectively.}   \\ 
\hline
$\alpha_F$ & \makecell[l]{Compressor's decision on whether or not to request feedback.} \\ 
\hline
$\sigma_S$ & \makecell[l]{Compressibility of the header source:
If $\sigma_S=0$ (not \\ compressor), IR or CO7 header must be used.} \\
\hline
$\sigma_D$ & \makecell[l]{State of decompressor.}  \\  
\hline
$\sigma_T$ & \makecell[l]{Transmission status.}  \\  
\hline
$\sigma_H$ & \makecell[l]{Channel condition: Transmission is more likely to be \\ successful when channel condition is better.}  \\  
\hline
$z_T$ &  \makecell[l]{Delayed observation of $\sigma_T$.} \\  
\hline
$z_D$ &  \makecell[l]{Feedback from the decompressor of its state, which has value \\ $-1$ if the compressor did not request a feedback.}  \\  
\hline
$z_H$ & \makecell[l]{Delayed observation of $\sigma_H$.} \\
\hline
\end{tabular}
\label{tab:1}
\end{table}


\section{POMDP Problem Formulation}\label{sec:pomdp}
In this section, we formulate the BD-ROHC as a partially observable \emph{Markov} decision process (POMDP). In the context of RL \cite{sutton2018reinforcement}, agent is the compressor, and the environment consists of the decompressor and the channels. Our goal is to find a policy for the compressor in the framework of POMDP. The BD-ROHC model is summarized in Fig. \ref{fig:rohcmodel}.

A POMDP is a $7$-tuple $(\mathcal{S}, \mathcal{A}, \mathcal{T}, \mathcal{R}, \mathcal{Z}, \mathcal{O}, \gamma)$, in which: $\mathcal{S}$ is the set of states; $\mathcal{A}$ is the set of actions; $\mathcal{T}: \mathcal{S}\times \mathcal{A}\rightarrow \Delta \mathcal{S}, (\sigma,\alpha) \mapsto \mathcal{T}(\cdot|\sigma,\alpha)$ is the transition dynamic, where $\Delta \mathcal{S}$ denotes the set of probability distributions defined on $\mathcal{S}$; $\mathcal{R}: \mathcal{S}\times \mathcal{A} \times \mathcal{S}\rightarrow \mathbb{R}_+, (\sigma,\alpha,\sigma')\mapsto \mathcal{R}(\sigma,\alpha,\sigma')$ is a reward function, where $\sigma'$ is the next state; $\mathcal{Z}$ is the sample space of the observations; $\mathcal{O}: \mathcal{S}\rightarrow \Delta \mathcal{Z}, \sigma \mapsto \mathcal{O}(\cdot|\sigma)$ is the conditional probability distribution of observation; $\gamma$ is the discounting factors. Let $d= \max(d_S,d_H,d_D)$, we now define the POMDP for our BD-ROHC system:

\begin{itemize}
\item The \textbf{state} variable $\sigma[t]\in \mathcal{S}$ is defined as 
\begin{gather}
\begin{aligned}
\sigma[t]=&(\alpha_C[t\!-\!1\!:\!t\!-\!d\!-\!1], \alpha_F[t\!-\!1\!:\!t\!-\!d\!-\!1],\\ 
& \sigma_S[t\!:\!t\!\!-\!d],  \sigma_D[t\!-\!d],  \sigma_T[t\!-\!d],  \sigma_H[t\!-\!d]) 
\end{aligned}
\end{gather}
Naturally, the set of states is $\mathcal{S}=\{0,1,2\}^{d+1}\times \{0,1\}^{d+1} \times \{0,1\}^{d+1} \times \{0,1,...,W+1\} \times \{0,1\} \times \mathcal{S}_H$. 
\item The \textbf{action} variable $\alpha[t]\in \mathcal{A}$ is defined as
\begin{gather}
\begin{aligned}
\alpha[t] = (\alpha_C[t],\alpha_F[t])
\end{aligned}
\end{gather}
with $\mathcal{A}=\{0,1,2\}\times \{0,1\}$.
\item The \textbf{partial observation} variable $z[t]\in \mathcal{Z}$ is defined as
\begin{gather}
z[t] = (z_T[t], z_H[t], z_D[t],  \sigma_S[t\!:\!t\!\!-\!d])    
\end{gather}
with $\mathcal{Z}=\{0,1\}\times \mathcal{Z}_H\times \{-1,0,1,...,W+1\}\times \{0,1\}^{d+1}$. Note that if the compressor receives feedback at $t$, then $z_D[t]= \sigma_D[t-d]$, otherwise we use $z_D[t]=-1$ to denote not receiving feedback.
\item The \textbf{observation probabilistic} $\mathcal{O}(z[t]|\sigma[t])$ is defined as
\begin{gather}
\begin{aligned}
\mathcal{O}&(z|\sigma[t]) \!= \mathcal{O}_T(\bar{z}_T| \sigma_T[t\!-\!d])\mathcal{O}_H(\bar{z}_H| \sigma_H[t\!-\!d])\\
&\times\mathbf{1}_{(\alpha_F[t-d]=0\wedge \bar{z}_D=-1)\vee (\alpha_F[t\!-\!d]=1\wedge \bar{z}_D= \sigma_D[t-d])} \\
&\times\mathbf{1}_{\bar{\sigma}_S= \sigma_S[t:t-d]}
\end{aligned}
\end{gather}
where $z=(\bar{z}_T, \bar{z}_H, \bar{z}_D, \bar{\sigma}_S)$, and $\mathbf{1}_{(\cdot)}$ is the indicator function returning $1$ if the condition in ``$(\cdot)$" is satisfied, returning $0$ otherwise.
\item The \textbf{transition dynamic} $\mathcal{T}(\sigma[t\!+\!1]|\sigma [t],\alpha[t])$ is defined as the follows
\begin{gather}
\begin{aligned}
&\mathcal{T}(\sigma|\sigma[t],\alpha[t])=   \mathcal{T}_T(\bar{\sigma}_T|\alpha_C[t\!-\!d], \sigma_H[t\!-\!d\!+\!1])\\
&\times \!\mathcal{T}_S(\bar{\sigma}_S| \sigma_S[t\!:\!t\!-\!d])\mathcal{T}_H(\bar{\sigma}_H| \sigma_H[t\!-\!d]) \\
&\times \! P_D(\bar{\sigma}_D| \sigma_T[t\!-\!d\!+\!1], \alpha_C[t\!-\!d],  \sigma_D[t\!-\!d]) \\
&\times \mathbf{1}_{\bar{\alpha}_C=\alpha_C[t:t-d]}\mathbf{1}_{\bar{\alpha}_F=\alpha_F[t:t-d]}
\end{aligned}
\end{gather}
where $s = (\bar{\alpha}_C, \bar{\alpha}_F, \bar{\sigma}_S, \bar{\sigma}_D, \bar{\sigma}_T, \bar{\sigma}_H)$.
\item The \textbf{reward} function $\mathcal{R}(\sigma[t], \alpha[t], \sigma[t\!+\!1])$ is defined as
\begin{gather}\label{eq:reward}
\mathcal{R}(\sigma[t], \alpha[t],\sigma[t\!+\!1]) \! = \! \!\frac{L\mathbf{1}_{ \sigma_D[t\!-\!d\!+\!1]\!=\!0}}{L+L_{\alpha_C[t-d]}} \! - \!\lambda \alpha_F[t\!-\!d\!-\!1]
\end{gather}
where $\lambda > 0$ is a constant. The first term on the right hand side of eq. (\ref{eq:reward}) accounts for the packet's data-efficiency. Recall that $L$ and $L_i$ are payload size and the header size corresponding to the action (header type) $\alpha_C = i$, respectively. When decoding fails, the the packet's data-efficiency is $0$. When decoding successes, i.e., $ \sigma_D = 0$, the packet's data-efficiency is $\frac{L}{L+L_{\alpha_C[t-d]}}$.  The second term penalizes the feedback since it introduces additional communication costs. The reward at the $t$-th slot is simply denoted as $r[t]  = \mathcal{R}(\sigma[t], \alpha[t], \sigma[t\!+\!1])$.
\end{itemize}

Note that in the POMDP, the agent (compressor) only has access to the partial observation $z[t]$ (of the state $\sigma[t]$) instead of $\sigma[t]$ itself. The agent has to rely on the history of observations $z[t\!:\!0]$ and actions $\alpha[t\!-\!1\!:\!0]$ to make decision $\alpha[t]$. We denote the deterministic policy as $\pi: \mathcal{F}\rightarrow \mathcal{A}$, where $\mathcal{F}$ is the history of the observations and actions, i.e., $\forall t, (z[t\!:\!0], \alpha[t\!-\!1\!:\!0]) \in \mathcal{F}$. Let $\alpha^\pi[t]=(\alpha_C^\pi[t],\alpha_F^\pi[t])$ be the action taken under policy $\pi$, then our goal is to find the policy $\pi$ maximizing the accumulated discounted reward:
\begin{gather}\label{eq:opt}
    \max_{\pi} \mathsf{E}\left[\sum_{t=0}^\infty \gamma^t\mathcal{R}(\sigma[t], \alpha^\pi[t], \sigma[t\!+\!1])\right].
\end{gather}

\begin{figure*}[htbp]
\centering
\includegraphics[width=0.65\linewidth]{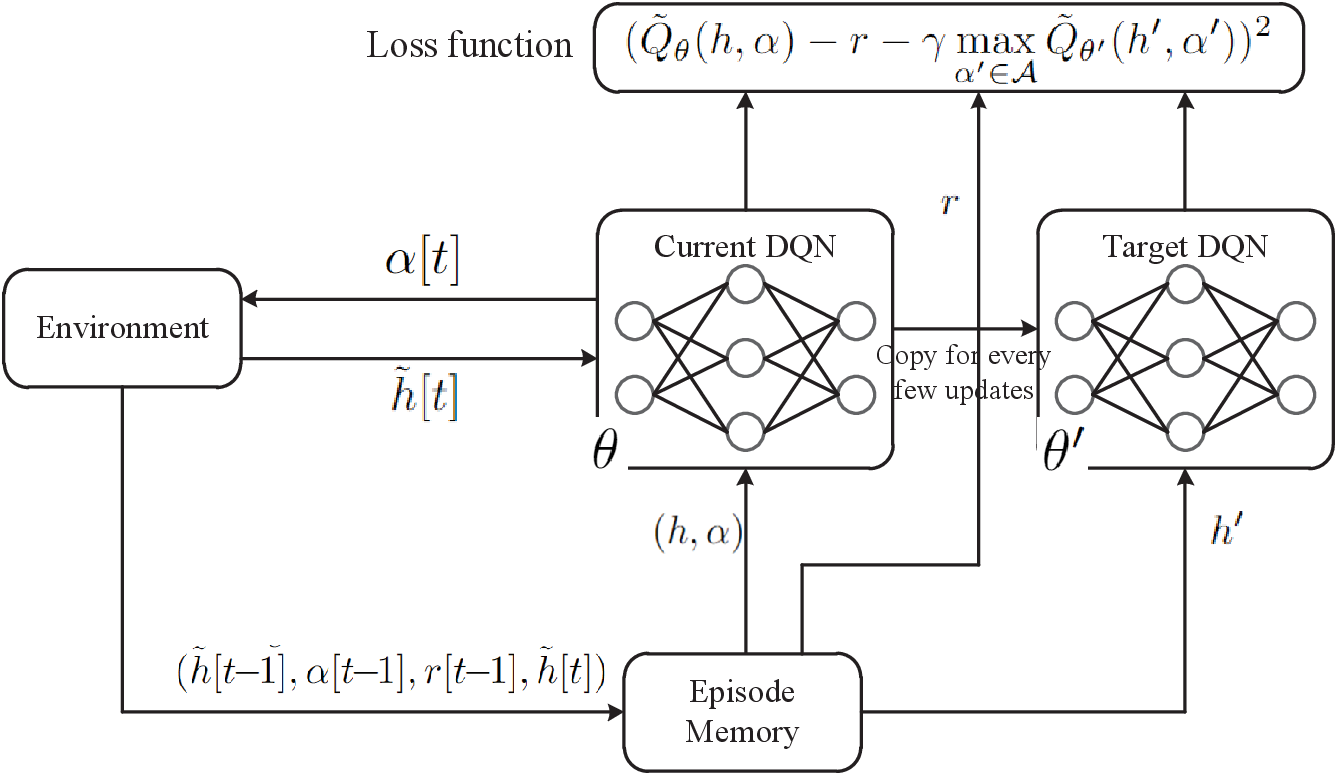}
\caption{The training paradigm of DDQN. The current DQN interacts with the environment (channel and decompressor) to collect experience. During the update of $\theta$, the target network $\theta'$ remains unchanged, and participate in calculating the target}
\label{fig:ddqn}
\end{figure*}

In our formulation,  the size of action space, state space and observation space are $|\mathcal{A}|=6$, $|\mathcal{S}| = (W+3)|\mathcal{S}_H|12^d$ and $|\mathcal{Z}| = 2(W+3)|\mathcal{Z}_H|2^d$, respectively. The complexity of  solving a POMDP with dynamic programming (DP) \cite{wenhao2} is exponential to $|\mathcal{A}| \times |\mathcal{Z}|$ and linear to $|\mathcal{S}|$, which becomes prohibitive when the scale of the problem becomes large, especially when $d$ is large.

\section{Compressor Policy with Deep Q-learning}\label{sec:dqn}
A prior work \cite{wenhao2} uses dynamic programming (DP) to solve the problem formulated in  (\ref{eq:opt}) which exhibits two shortcomings: 1) DP becomes computationally prohibitive when the scale of the problem becomes large due to large delay $d$ and complex channel models. 2) DP requires the knowledge of transition dynamic $\mathcal{T}$, which is often not available in many practical applications. As we shall see in this section, these two obstacles can be addressed by our proposed deep Q-learning methods.
\subsection{Deep Q-learning for MDP}
Before moving onto deep Q-learning for POMDP, it is helpful to review the Q-learning for MDP where states are available. The key quantity plays in Q-learning is the Q-function. In an MDP, the Q-function $Q^\kappa: \mathcal{S}\times \mathcal{A}\rightarrow \mathbb{R}$ with a deterministic policy $\kappa: \mathcal{S}\rightarrow \mathcal{A}$ is defined as 
\begin{gather}\label{eq:qmdp}
Q^\kappa(\sigma,\alpha) = \mathsf{E}\left[ \sum_{k=0}^\infty  \gamma^k r[t+k] \Bigm\vert   \sigma[t]=\sigma, \alpha[t]=\alpha, \kappa \right].  
\end{gather}
The Q-value (value of Q-function) can be explained as the expected discounted accumulated rewards starting from state $s$ taking action $a$ under policy $\kappa$. The goal of Q-learning is to find the optimal Q-function (corresponding to the optimal policy $\kappa$) in a sense that $Q^\kappa(\sigma,\alpha)$ is maximized for every $\sigma$ and $\alpha$. For most cases, there is no closed-form formulation of Q-function in terms of $\kappa$. 

According to \cite{sutton2018reinforcement}, the optimal policy satisfies
\begin{gather}\label{eq:kmdp}
\kappa^* (\sigma) = \arg\max_{\alpha\in\mathcal{A}} Q^{\kappa^*}(\sigma,\alpha).
\end{gather}
Applying the recursion of  Q-function, and substituting eq. (\ref{eq:kmdp}) into eq. (\ref{eq:qmdp}), we have 
\begin{gather}\label{eq:qmdp2}
Q^{\kappa^*}(\sigma,\alpha) \!= \!\sum_{\sigma'\in \mathcal{S}}\!\!\mathcal{T}(\sigma'|\sigma,\alpha)\left(\mathcal{R}(\sigma,\alpha,\sigma')\! + \!\gamma Q^{\kappa^*}(\sigma'\!,\kappa^*(\sigma')) \right) \nonumber \\
=\! \!\sum_{\sigma'\in \mathcal{S}}\!\!\mathcal{T}(\sigma'|\sigma,\alpha)\!\!\left(\!\mathcal{R}(\sigma,\alpha,\sigma')\! + \! \gamma \max_{\alpha'\in \mathcal{A}}Q^{\kappa^*}(\sigma'\!,\alpha') \!\right)\!.
\end{gather}
which is the \emph{Bellman} equation where $Q(\sigma,\alpha)$ can be viewed as the unknown table to be solved with finite dimension $|\mathcal{S}|\times |\mathcal{A}|$.
It has been proved that the Q-function (or the policy) is optimal if and only if the \emph{Bellman} equation is satisfied \cite{sutton2018reinforcement}. To solve the equation, the Q-function can be updated as follows
\begin{gather}\label{eq:iter}
Q(\sigma,\alpha) \leftarrow r+\gamma \max_{\alpha'\in \mathcal{A}}Q(\sigma',\alpha')
\end{gather}
where $(\sigma,\alpha,r,\sigma')$ denote the state, action, reward and the next state, respectively, sampled from trajectories (instantiations of the MDP). Eq. (\ref{eq:iter}) is guaranteed to converge to the optimal $Q^{\kappa^*}(\sigma,\alpha)$ according to \cite{melo2001convergence}.

When the state space and the action space both become large (infinite if $\sigma$ and $\alpha$ are continuous), we can use a neural network (NN) to represent the Q-function $Q_\theta(\sigma,\alpha)$, where $\theta$ are the weights of the NN. In this case, for each $(\sigma,\alpha,r,\sigma')$, the sample loss can be written as
\begin{gather}
    g(\sigma,\alpha,r,\sigma';\theta) = (Q_\theta(\sigma,\alpha) \!- \!r - \! \gamma\max_{\alpha'\!\in \mathcal{A}}Q_\theta(\sigma'\!,\alpha') )^2.
\end{gather}
The update of $\theta$ follows the mini-batch stochastic gradient descent (SGD): $\theta \leftarrow \theta - \eta\nabla g(\sigma,\alpha,r,\sigma')$, where $\eta$ is the learning rate. Note that, deep Q-learning is not guaranteed to converge with non-linear NN model. Its convergence is still an open problem under exploration.

\subsection{Double Deep Q-learning Implementation for the Compressor Policy in POMDP}
In the POMDP, state $\sigma$ is not available, consequently we replace $Q_\theta(\sigma,\alpha)$ with $\Tilde{Q}_\theta(h,\alpha)$, where $h\in \mathcal{F}$ is the history of partial observations and actions. Despite lacking theoretical guarantee, it works well in many POMDP applications.

\subsubsection{Deep Q-network}
Let $f_\theta(\cdot)$ be the function represented by the deep Q-network (DQN). It takes truncated history 
\begin{gather}
\tilde{h}[t]\triangleq (z[t\!:\!t\!-\!d\!-\!d_0], \alpha[t\!-\!1\!:\!t\!-\!d\!-\!d_0]) 
\end{gather}
as the inputs and has a $|\mathcal{A}|$-dimension output, in which the $\alpha$-th entry $f^\alpha_\theta (\tilde{h}[t]) = \Tilde{Q}_\theta(\tilde{h}[t],\alpha)$. The constant integer $d_0>0$ is used to adjust the history window size. Technically, all the histories of observations are informative to decision making. In DQN, we only keep the latest observations and actions, since the early ones have smaller impact on the current state.

\subsubsection{Double deep Q-learning}

As mentioned previously, deep Q-learning often suffers from instability of convergence resulting in unsatisfactory performances. In this work, double DQN (DDQN) is adopted to mitigate this issue, whose block diagram is depicted in Fig. \ref{fig:ddqn}. In DDQN, there are two networks, namely current DQN and target DQN, parameterised with $\theta$ and $\theta'$, respectively. The current DQN interacts with the environment (decompressor, channel and header source) to collect experience (finally stored in the episode memory $\mathcal{M}$). During the update of $\theta$, the target network $\theta'$ remains unchanged, and participate in calculating the target $r + \gamma\max_{\alpha'\in \mathcal{A}}\tilde{Q}_{\theta'}(h',\alpha')$ and the following sample loss.
\begin{gather}
\begin{aligned}
&\tilde{g}(h,\alpha,r,h';\theta,\theta') = (\tilde{Q}_\theta(h,\alpha)  - r - \gamma \max_{\alpha'\in \mathcal{A}}\tilde{Q}_{\theta'}(h',\alpha') )^2.    
\end{aligned}
\end{gather}
After several updates of the current network $\theta$, DDQN updates the target network $\theta'$ by copying $\theta$. The detailed algorithm for DDQN is shown in Algorithm \ref{alg:2p}. 

\begin{algorithm}[htbp]
\caption{Training DDQN Compressor for BD-ROHC}
\begin{algorithmic}[1]
\STATE{\textbf{Initialization:} Initialize the current and target network with $\theta$ and $\theta'$, respectively. Random exploration probability $\epsilon = 1$.}
\FOR{$j = 0,1,..., M$}
\FOR{$t = 0,1,..., T$}
\STATE{\textbf{Compressor:}}
\STATE{Obtain $ \sigma_S[t]$ from the header source}
\STATE{Obtain $z_T[t]$, $z_H[t]$ from the trans-layer information based on $ \sigma_T[t\!-\!d]$ and $ \sigma_H[t\!-\!d]$}
\STATE{Obtain $z_D[t]= \sigma_D[t\!-\!d]$ if $\alpha_F[t\!-\!d\!-\!1]=1$,  \\ $z_D[t]=-1$ if $\alpha_F[t\!-\!d\!-\!1]=0$}
\IF{$\epsilon < rand()$}
\STATE{$\alpha[t]\sim Uniform(\mathcal{A})$}
\ELSE
\STATE{Collect the history of $ \sigma_S$, $z_T$, $z_H$, $\alpha_C$, $\alpha_F$ up to the $(t\!-\!d\!-d_0)$-th slot, and concatenate them as $\tilde{h}[t]$}
\STATE{$\alpha[t] = \max_{a\in\mathcal{A}} \tilde{Q}_{\theta}(\tilde{h}[t],a)$}
\ENDIF
\STATE{$(\alpha_C[t], \alpha_F[t])=\alpha[t]$}
\STATE{Transmit $\alpha_C[t]$}
\STATE{Poll feedback according to $\alpha_F[t]$}
\STATE{\textbf{Decompressor:}}
\STATE{Receive $\alpha_C[t\!-\!d]$ with channel quality $ \sigma_H[t\!-\!d\!+\!1]$ and transmission status $ \sigma_T[t\!-\!d\!+\!1]$}
\STATE{Update state $ \sigma_D[t\!-\!d\!+\!1]$ based on $\alpha_C[t\!-\!d]$, $ \sigma_T[t\!-\!d\!+\!1]$ and $ \sigma_D[t\!-\!d\!+\!1]$ }
\STATE{Feedback $ \sigma_D[t\!-\!d\!+\!1]$ if $\alpha_F[t\!-\!d\!-\!1]=1$}
\STATE{\textbf{Episode Memory:}}
\STATE{Collect the reward $r[t] \!=\!\frac{L\mathbf{1}_{ \sigma_D[t\!-\!d\!+\!1]\!=\!0}}{L+L_{\alpha_C[t-d]}} \! - \!\lambda \alpha_F[t\!-\!d\!-\!1]$}
\STATE{Update memory $\mathcal{M}$ with $(\tilde{h}[t\!-\!1],\alpha[t\!-\!1], r[t\!-\!1], \tilde{h}[t])$}
\ENDFOR
\STATE{\textbf{Training:}}
\FOR{$k = 0,1,...$}
\STATE{$(\tilde{h},\alpha,r,\tilde{h}')\sim \mathcal{M}$}
\STATE{$\theta\leftarrow\theta-\eta \nabla \tilde{g}(h,\alpha,r,h';\theta,\theta')$}
\ENDFOR
\STATE{$\theta'\leftarrow\theta$}
\STATE{$\epsilon \leftarrow \epsilon\gamma_{\epsilon}$}
\ENDFOR
\end{algorithmic}
\label{alg:2p}
\end{algorithm}

We adopt nn $\epsilon$-greedy strategy such that the agent can explore the action spaces in early training stages. At the beginning, the agent has large probability $\epsilon$ close to $1$ to take random actions. The value of  $\epsilon$ decays with rate $\gamma_\epsilon$. After a duration, the agent will follow the current policy with high probability. Note that there is a $d$-slot delay between the compressor and the decompressor. For this reason, while the compressor transmits $\alpha_C[t]$, the decompressor just receives $\alpha_C[t-d]$. We denote
the episode memory as $\mathcal{M}$, which is essentially a first-in-first-out (FIFO) with constant size.

It is worth mentioning that the proposed DDNQ method finds the policy with trajectory samples obtained from interacting with the environments, and does not require any prior knowledge of the transition dynamic or observation probabilistic. In addition, the proposed method can handle state, action and observations from large finite or even continuous spaces.

\section{Experimental Results} \label{exp1}
We start with the general settings of the experiments.
During the training, the number of episode is set as $M=3,000$, and $T=10,000$ packets are transmitted within each episode. Both the current and target DQN are $4$-layer dense NN, in which the hidden layer has width $2,048$.

\begin{itemize}
    \item \textbf{Headers}: The length of IR header ($\alpha_C=0$), CO7 header ($\alpha_C=1$) and CO3 header ($\alpha_C=2$) are set as $L_0=60$, $L_1=15$ and $L_2=1$, respectively. The header source $ \sigma_S[t]$ is modelled as a \emph{Markov} process with order $d_S=1$ and transition dynamic $\mathcal{T}_S( \sigma_S[t]| \sigma_S[t\!-\!1\!])$. In the simulation, $\mathcal{T}_S( \sigma_S[t]=1| \sigma_S[t\!-\!1\!]=0)=1$ and $\mathcal{T}_S( \sigma_S[t]=0| \sigma_S[t\!-\!1\!]=1)=0.1$.
    \item \textbf{Decompressor}: The decompressor follows the model in Sec. \ref{sec:decomp} with the maximum number of allowed consecutive decoding failures as $W=5$.
    \item \textbf{Evaluation metric}: We evaluate ``transmission efficiency", which is defined as total length of corrected received payloads over total length of the packet:
        \begin{gather}\label{eq:eff}
        \text{transmission efficiency} = \frac{\sum_{t=1}^T\mathbf{1}_{ \sigma_D[t+1]=1}L}{\sum_{t=1}^T(L+L_{\alpha_C[t]})}.
        \end{gather}
    We also evaluate ``feedback rate", which is defined as the number of feedback requests over the total number of transmitted packets:
        \begin{gather}
        \text{feedback rate} = \frac{\sum_{t=1}^T \alpha_F[t]}{T}.
        \end{gather}
    \item \textbf{Benchmarks}: Given existing works can not handle situations with long feedback delays and unavailable transition dynamics, we propose the following ``keep transmitting" (KT) algorithm as the benchmark. KT requests feedback randomly with certain probability (such that feedback rate is controlled), and chooses header based on the latest feedback from the decompressor. KT keeps using the same header based on the last feedback until new feedback is received.
\end{itemize}

\subsection{Results under Gilbert-Elliot Channel Model}\label{sec:ge}
First, we consider the well known Gilbert-Elliot channel model \cite{rohc2}. It is parameterized with the average duration of ``bad" states $l_B$ and the probability of ``bad" state $\epsilon_B$. The ``bad" state is represented as $ \sigma_H[t]=0$ and ``good'' state is represented as $ \sigma_H[t]=1$. The transition dynamic $\mathcal{T}_H( \sigma_H[t]| \sigma_H[t\!-\!1])$ has values $\mathcal{T}_H( \sigma_H[t]=1| \sigma_H[t\!-\!1]=0)=\frac{1/l_B}{1/\epsilon_B -1}$ and $\mathcal{T}_H( \sigma_H[t]=0| \sigma_H[t\!-\!1]=1)=1/l_B$. The transmission status has dynamic $\mathcal{T}_T( \sigma_T[t]=1| \sigma_H[t]= 1) = \beta_1$ and $\mathcal{T}_T( \sigma_T[t]=1| \sigma_H[t]= 0) = \beta_0$. Through this subsection, we set $l_B=5$.

The trans-layer information is summarized as the estimates $z_H[t]$ and $z_T[t]$ through model $\mathcal{O}_H(z_H[t]| \sigma_H[t\!-\!d_D])$ and $\mathcal{O}_T(z_T[t]| \sigma_T[t\!-\!d_D])$, respectively. In our simulation, $\mathcal{O}_H(z_H[t]= \sigma_H[t\!-\!d_D]| \sigma_H[t\!-\!d_D])=1-\epsilon_H$, and $\mathcal{O}_T(z_T[t]= \sigma_T[t\!-\!d_D]| \sigma_T[t\!-\!d_D])=1-\epsilon_T$, where $\epsilon_T$ and $\epsilon_H$ denote the estimation error probability of transmission status and channel conditions, respectively. It is worth to clarify that channel condition $ \sigma_H[t]$ indicates how good the channel is at the $t$-th slot, while $\epsilon_B$ parameterize the probabilistic model of $ \sigma_H[t]$.

\begin{figure}[htbp]
\centering
\includegraphics[width=\linewidth]{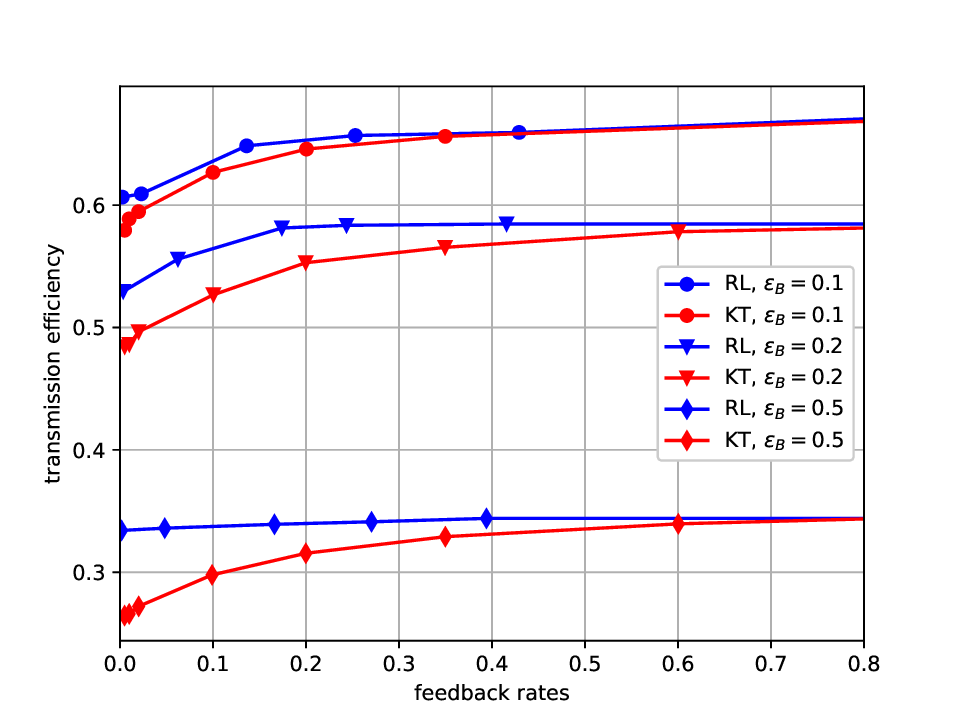}
\caption{Performance of the proposed RL and KT under different channel quality parameter $\epsilon_B$. The proposed RL outperforms KT, and the performance gap becomes more obvious as feedback rate decrease. (Gilbert-Elliot model)}
\label{fig:epsB}
\end{figure}

Fig. \ref{fig:epsB} shows the results of the proposed RL and KT under different channel quality $\epsilon_B$. It can be observed from the figure that the performance gap between the two methods becomes more obvious as feedback rate decreases. Since the problem under low feedback rate is more challenging, the carefully designed RL method can show more advantages without surprise. The proposed RL has better performance than KT with all different channel qualities. In the experiment, $\epsilon_T = 0.1$, $\epsilon_H=0.1$, $d=4$, $\beta_1 = 0.9$, $\beta_0 = 0.1$ and $L=20$.


\begin{figure}[htbp]
\centering
\includegraphics[width=\linewidth]{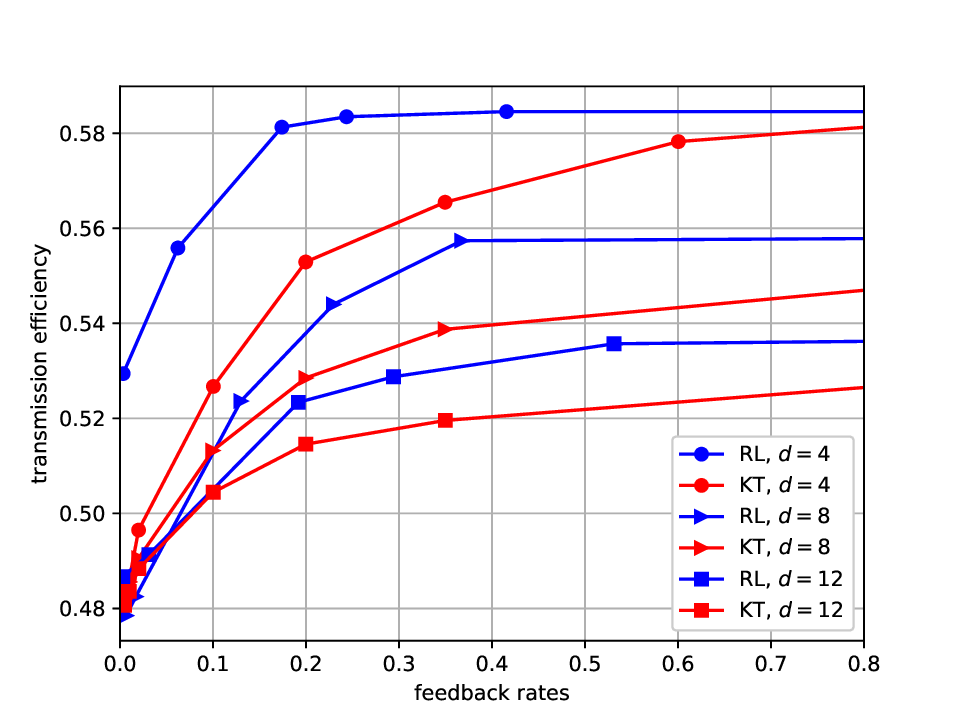}
\caption{Performance of the proposed RL and KT under different feedback delay $d$.  The proposed RL method outperforms KT, and both methods performs worse as the feedback delay $d$ becomes larger since the feedback becomes less informative. (Gilbert-Elliot model)}
\label{fig:d}
\end{figure}

Fig. \ref{fig:d} shows the performance of both methods under different feedback delay $d$. It can be observed that the proposed RL method outperforms KT, and both methods performs worse as the feedback delay $d$ becomes larger, since the feedback becomes less informative. In the experiment, $\epsilon_B=0.2$, $\epsilon_T = 0.1$, $\epsilon_H=0.1$, $\beta_1 = 0.9$, $\beta_0 = 0.1$ and $L=20$.

\begin{figure}[htbp]
\centering
\includegraphics[width=\linewidth]{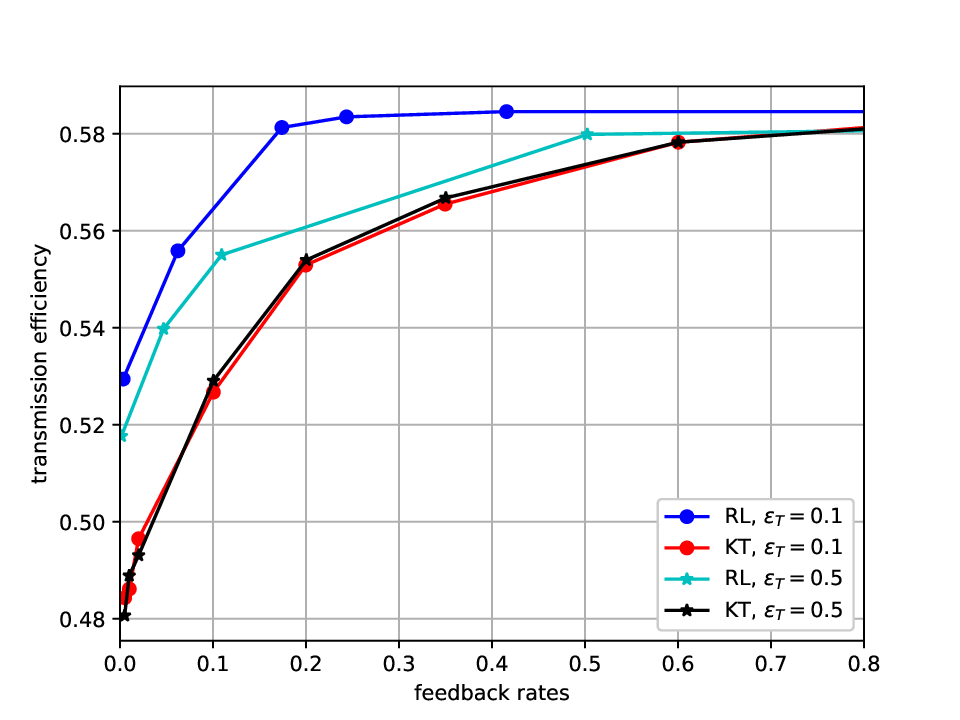}
\caption{Performance of the proposed RL and KT under different transmission status estimation error probability $\epsilon_T$. (Gilbert-Elliot model)}
\label{fig:epsT}
\end{figure}

\begin{figure}[htbp]
\centering
\includegraphics[width=\linewidth]{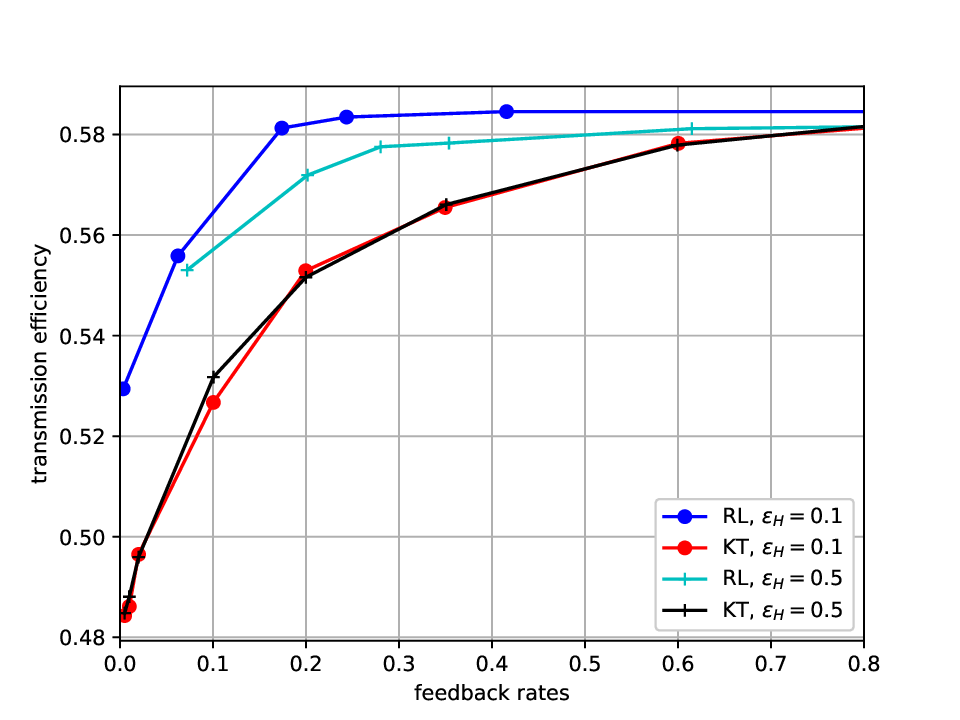}
\caption{Performance of the proposed RL and KT under different channel condition estimation error probability $\epsilon_H$. (Gilbert-Elliot model)}
\label{fig:epsH}
\end{figure}

Fig. \ref{fig:epsT} shows the performances of both methods with different transmission status estimation error probability $\epsilon_T$. It can be observed that the proposed RL method has better performance when $\epsilon_T$ is smaller. The performance of KT does not change along with $\epsilon_T$, because it can not use trans-layer information without knowledge of the model. In the experiment, $\epsilon_B=0.2$, $\epsilon_H=0.1$, $d=4$ and $L=20$. In Fig. \ref{fig:epsH}, we made observations similar to  Fig. \ref{fig:epsT}. In the experiment, $\epsilon_B=0.2$, $\epsilon_T=0.1$, $d=4$, $\beta_1 = 0.9$, $\beta_0 = 0.1$ and $L=20$.

\begin{figure}[htbp]
\centering
\includegraphics[width=\linewidth]{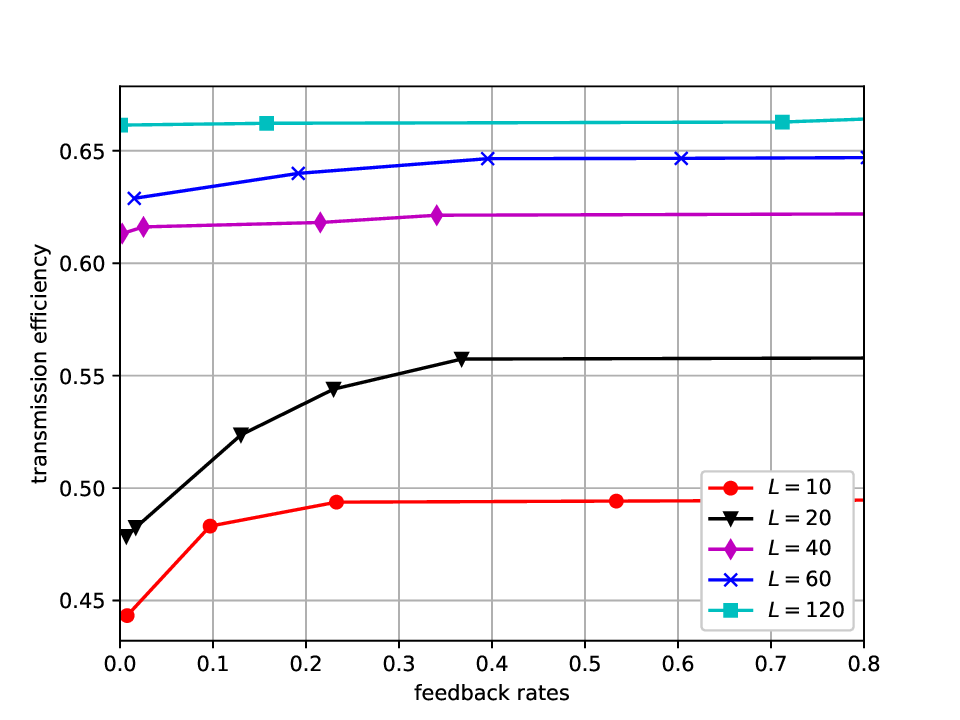}
\caption{Performance of the proposed RL under different payload size $L$. Transmission efficiency is higher with larger payload. (Gilbert-Elliot model)}
\label{fig:L}
\end{figure}

Fig. \ref{fig:L} shows the impact of payload size $L$ on the transmission efficiency. It can be observed from the figure that larger payload size results in a high transmission efficiency. Notice that the payload size is not related to any transition dynamic and observation probabilistic. For any instantiation of the process, increase $L$ always results in high transmission efficiency. The limit of the transmission efficiency is $\sum_{t=1}^T\mathbf{1}_{ \sigma_D[t+1]=1}/T$ according to eq. (\ref{eq:eff}). In the experiment, $\epsilon_B=0.2$, $\epsilon_T=0.1$, $\epsilon_H=0.1$, $\beta_1 = 0.9$, $\beta_0 = 0.1$ and $d=4$.

\begin{figure}[htbp]
\centering
\includegraphics[width=\linewidth]{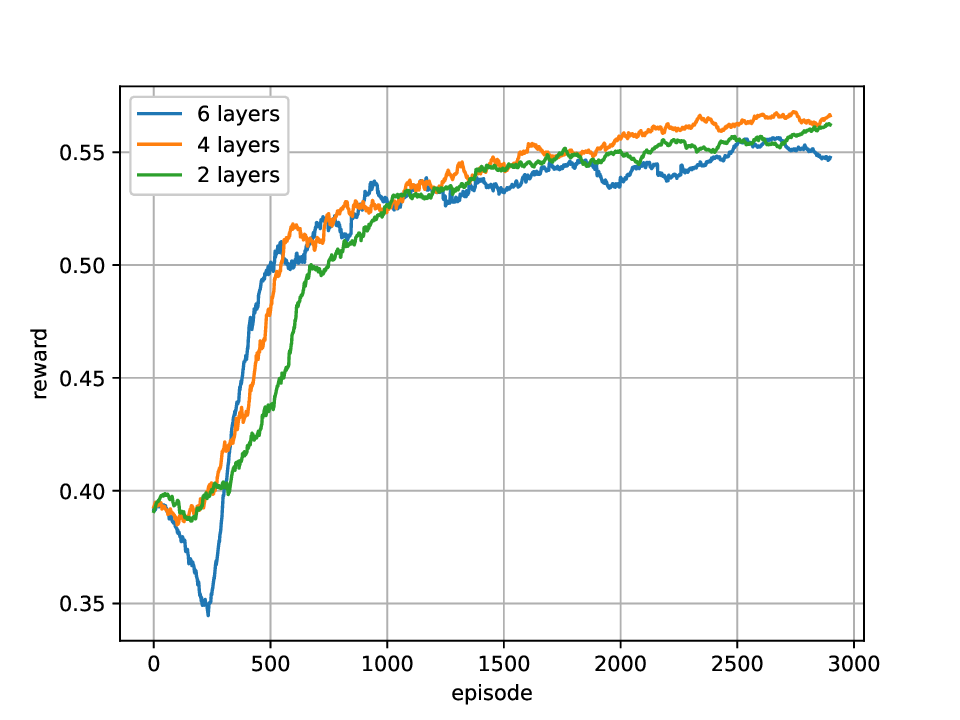}
\caption{Convergence of the training process with different number of layers. The curves are smoothed with a window of size $50$. (Gilbert-Elliot model)}
\label{fig:lay}
\end{figure}

Fig. \ref{fig:lay} shows performance of the proposed RL method with DQN with different depths. From the figure, we can observe that the DQN with $4$ layers perform better than the one with $2$ layers, as the later one is too simple thus lack of representation capability. However, when the DQN has $6$ layers, the performance becomes worse surprisingly. It is likely that the training of DQN becomes more unstable due to the increasing sensitivity resulting from deeper models.

\begin{figure}[htbp]
\centering
\includegraphics[width=\linewidth]{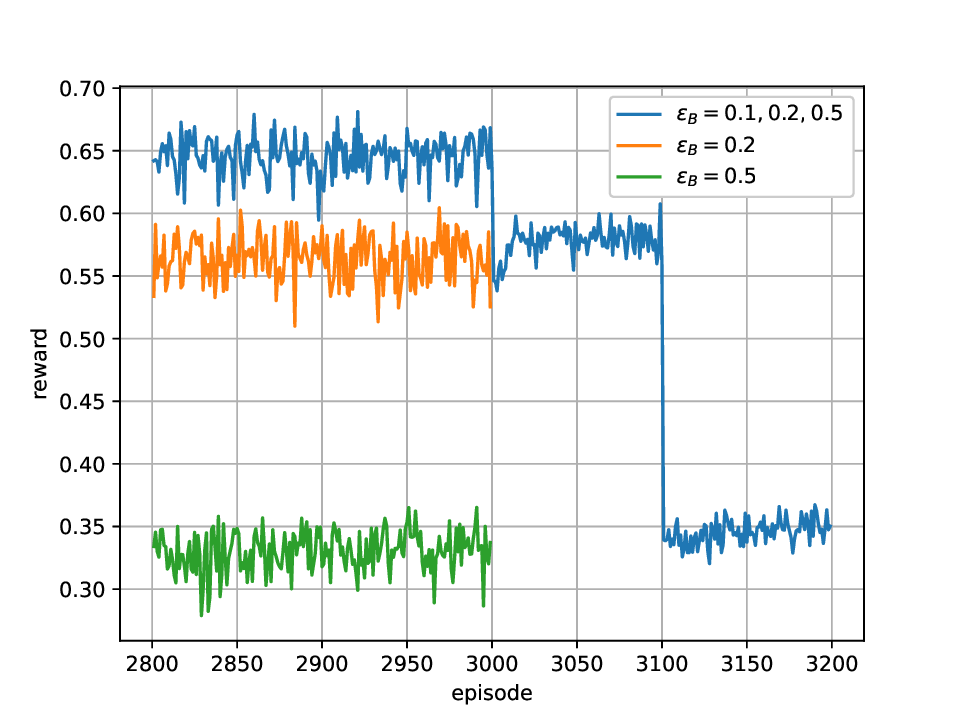}
\caption{Rewards during training with varying environments. In the experiment, $\epsilon_B=0.1$ before the $3,000$-th episode, and changes to $0.2$ and $0.5$ at the $3,001$-th and $3,101$-th episode, respectively. (Gilbert-Elliot model)}
\label{fig:adapt}
\end{figure}

Fig. \ref{fig:adapt} demonstrates how fast the proposed RL can adapt to new environment. The orange and green curves show the rewards from the $2,800$-th to the $3,000$-th episode during the training with $\epsilon_B=0.2$ and $\epsilon_B=0.5$, respectively. The blue curve shows the reward with varying $\epsilon_B$. Specifically, $\epsilon_B=0.1$ before the $3,000$-th episode, $\epsilon_B=0.2$ between the $3,001$-th episode and $3,100$-th episode, $\epsilon_B=0.5$ after the $3,101$-th episode. From the figure we observe that the proposed DQN can adopt to new environment quickly. The experiment setting is same to Fig. \ref{fig:epsB}.

\begin{figure}[htbp]
\centering
\includegraphics[width=\linewidth]{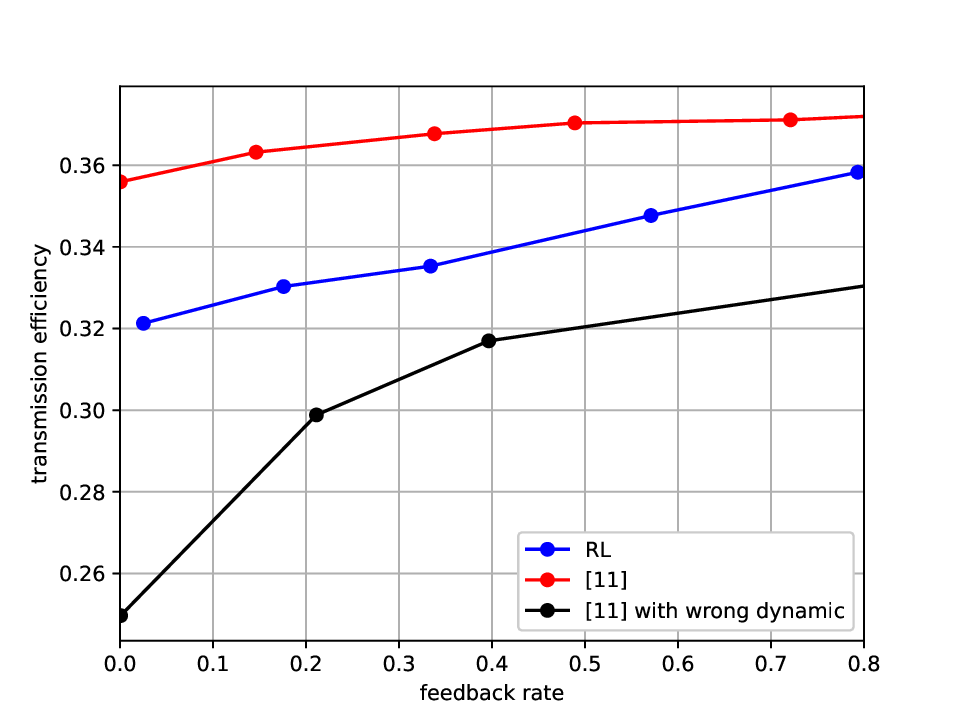}
\caption{Compared with \cite{wenhao2} with and without prior knowledge of model dynamics. (Gilbert-Elliot model)}
\label{fig:myopic}
\end{figure}

Fig. \ref{fig:myopic} shows a comparison between \cite{wenhao2} and the proposed RL. We can observe from the figure that \cite{wenhao2} outperforms the proposed RL with accurate knowledge of model dynamics. However, when the knowledge is inaccurate, it performs worse than the proposed RL. In the experiment, $\epsilon_B = 0.5$, $\epsilon_T = 0.4$, $\epsilon_H=0.4$, $d=4$, $\beta_1 = 0.7$, $\beta_0 = 0.3$ and $L=20$.

\subsection{Results under Hidden Markov Channel Model}
We now apply a hidden \emph{Markov} channel model \cite{hmm} to test the performance of the proposed RL and KT methods. The model starts from the physical layer wireless channel with \emph{Rayleigh} model,
\begin{gather}
     \sigma_H[t] = \sqrt{A_I[t]^2+A_Q[t]^2}
\end{gather}
where $A_I[t]$ and $A_Q[t]$ are in-phase and quadrature components of the channel, respectively. $A_I[t]$ and $A_Q[t]$ are independent $d_H$- order \emph{Markov} \emph{Gaussian} process. The transition dynamic is described by a $d_H\times d_H$ covariance matrix whose $(i,j)$-th entry is $\rho^{|i-j|}$. Here $\rho$ is a parameter to adjust the correlation of consecutive samples. The transmission status $ \sigma_T[t]$ can be expressed as
\begin{gather}
 \sigma_T[t] = \mathbf{1}_{P_TA[t]>U[t]}    
\end{gather}
where $P_T$ is transmitting power, and $U[t]$ servers as a threshold follows standard \emph{Gaussian} distribution independently at every time slot. The observation $z_H[t]$ is 
\begin{gather}
z_H[t] =  \sigma_H[t-d]+n_H[t-d]
\end{gather}
where $n_H[t-d]$ is an additive white \emph{Gaussian} noise (AWGN) with zero-mean and $\omega^2_H$-variance. We assume $z_H$ results from channel estimation and channel reciprocity. The observation $z_T$ is defined the same way with Gilbert-Elliot in Sec. \ref{sec:ge}. Through this subsection, we set $d_H=4$, $d=8$ and $L=20$.

\begin{figure}[htbp]
\centering
\includegraphics[width=\linewidth]{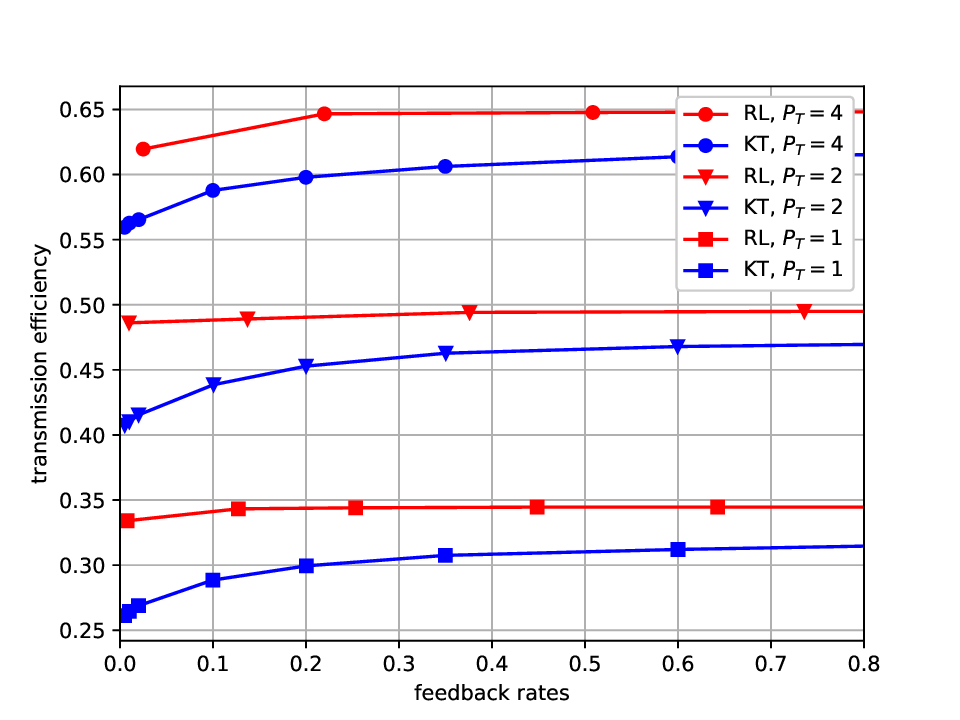}
\caption{Performance of the proposed RL and KT under different transmitting power $P_T$. (Hidden \emph{Markov} model)}
\label{fig:PT}
\end{figure}

Fig. \ref{fig:PT} shows the performances of both the proposed RL and KT methods with different transmitting power $P_T$. From the figure we observe that the propose RL methods outperforms KT in all cases, and their performance gap is relatively larger when feedback rate is lower. In the experiment, we set $\rho=0.5$ and $\omega_H^2=1$.

\begin{figure}[htbp]
\centering
\includegraphics[width=\linewidth]{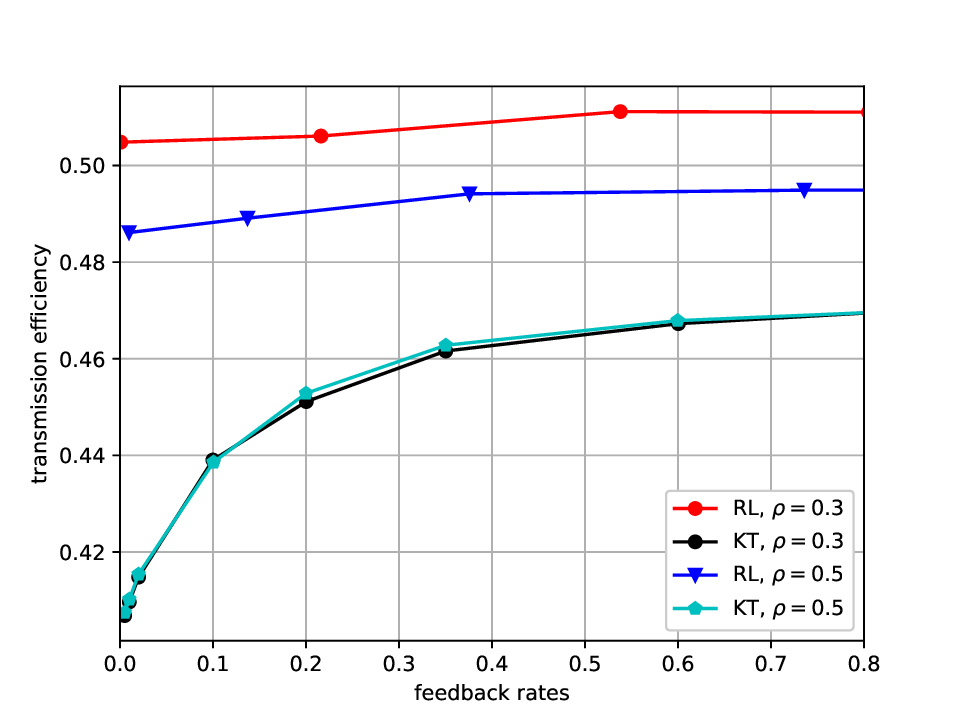}
\caption{Performance of the proposed RL and KT under different channel correlation $\rho$. (Hidden \emph{Markov} model)}
\label{fig:rho}
\end{figure}

Fig. \ref{fig:rho} shows the performances of both the proposed RL and KT methods with different channel correlation $\rho$. From the figure we observe that the performance of RL is better than KT in call cases, but degrades when $\rho$ becomes larger.  In the experiment, we set $P_T=2$ and $\omega_H^2=1$.

\begin{figure}[htbp]
\centering
\includegraphics[width=\linewidth]{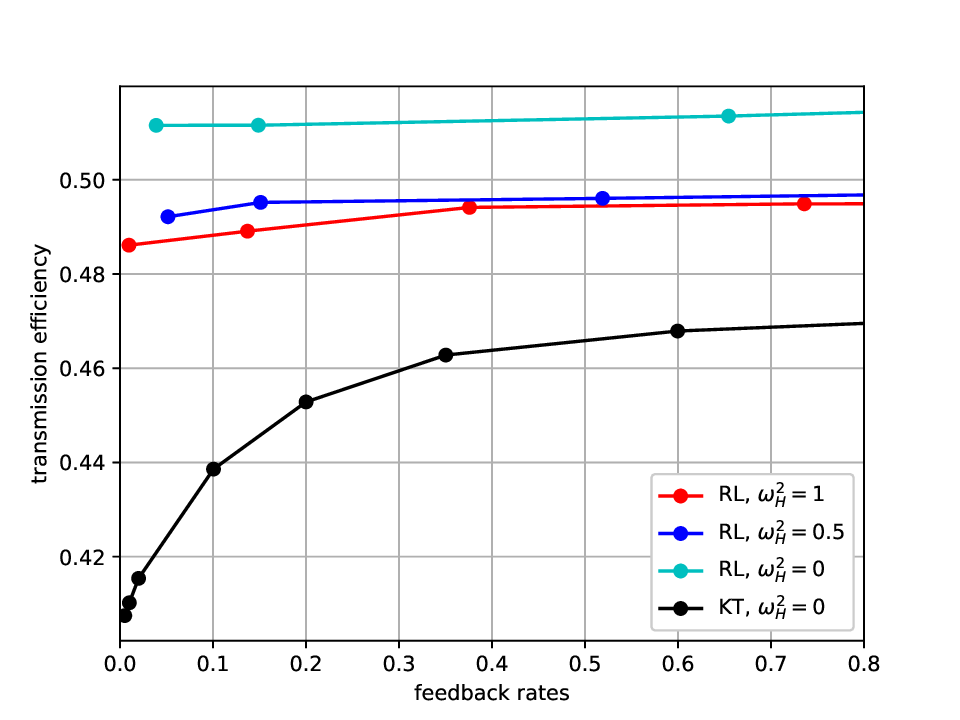}
\caption{Performance of the proposed RL and KT under different level of channel condition observation noise $\omega_H^2$. (Hidden \emph{Markov} model)}
\label{fig:omg}
\end{figure}

Fig. \ref{fig:omg} shows the impact of the variance of channel condition observation noise $\omega_H^2$ on the transmission efficiency. Without surprise, smaller observation noise results in better performance. It can be observed that 
the proposed RL method performs better than KT in all cases. In the experiment, we set $P_T=2$ and $\rho=0.5$.

\section{Conclusion}\label{sec:conl}
Existing works on bi-directional robust header compression (BD-ROHC) with dynamic programming (DP) are difficult to implement for large scale system due to prohibitive computational complexity. Moreover, dynamic programming ROHC controls rely on prior knowledge of the underlying model parameters, which is often unavailable practically. In this paper, we propose a novel RL framework which addresses these issues at the same time. We adopt a double deep Q-network (DDQN) framework, whose input dimension is scalable to the system model. Our training of the DDQN relies on
information obtained from interacting with channel and compressor, which can adaptively learn and acquire useful knowledge of the model dynamics implicitly. Experimental results demonstrate strong and robust performance of our proposed paradigm for different system models. Future work may consider more complex environment with multi-agent reinforcement learning.

\bibliographystyle{IEEEtran}
\begin{footnotesize}
\bibliography{mybib}
\end{footnotesize}

\end{document}